\documentclass[twocolumn,aps,amsmath,english,superscriptaddress,longbibliography,notitlepage]{revtex4-1}
\usepackage{graphicx,natbib,subfigure,tabularx,epsfig,longtable,amsfonts,rotating,mathrsfs}
\usepackage{color,babel,currfile,subfigure,hyperref}
\usepackage[normalem]{ulem}
\usepackage{float}
\usepackage{array}
\begin{document}
\title{Topological Phase Transition of A Non-Hermitian Crosslinked Chain}
\date{\today}

\author{X. L. Zhao}
\affiliation{Graduate School of China Academy of Engineering Physics, China\\}
\affiliation{Center for Quantum Sciences and School of Physics, Northeast Normal University, Changchun 
130024, China\\}

\author{L. B. Chen}
\affiliation{Quantum Physics Laboratory,  School of Science, Qingdao Technological University, Qingdao, 
266033, China\\}

\author{L. B. Fu} \thanks{ E-mail: lbfu@gscaep.ac.cn}
\affiliation{Graduate School of China Academy of Engineering Physics, China\\}

\author{X. X. Yi} \thanks{ E-mail: yixx@nenu.edu.cn}
\affiliation{Center for Quantum Sciences and School of Physics, Northeast Normal University, Changchun 
130024, China\\}
\begin{abstract}
Non-Hermiticity enriches the contents of topological classification of matter including exceptional points, 
bulk-edge correspondence and skin effect. Gain and loss can be described by imaginary diagonal 
elements in Hamiltonians and the topological phase transition for a crosslinked chain in the presence 
of such non-Hermiticity is investigated in this work. We obtain the phase diagram in term of a winding 
number analytically. The boundaries of the phases coincide with the surfaces of exceptional points in the 
parameter space. The topologically original edge states locating mainly at the joints between domains 
of different phases hold on even for the long chain. The non-Hermitian topological feature can also be 
reflected by vortex structures in the vector fields of complex eigenenergies and expected values of Pauli 
matrices or the trajectories of these quantities. This model can be implemented in coupled waveguides 
or photonic crystals. And the edge states are immune to various kinds of disorders until the topological 
phase transition occurs. This work benefits our insight into the influence of gain and loss on the 
topological phase of matter.
\end{abstract}
\maketitle
\section{Introduction}        
In the Hermitian realm, topological classification of matter benefits our understanding of condensed matter 
physics based on topological invariants in a global manner~\cite{RMP823045,RMP831057}. Non-Hermiticity 
can arise as asymmetric hopping or gain and loss in physical systems~\cite{PRL702273,PRL121213902,
NP1411,NC10297}. Extending topological classification theory for non-Hermitian systems may bring intriguing 
contents associated with some peculiar properties related to complex energy spectra, topological invariants 
and edge states, such as exceptional points\cite{PRE61929,CJP541039,PRL86787,PNAS1136845,PRL104153601,Nature525354,Nature526554,
arxiv180809541}, unusual bulk-edge correspondence \cite{JPAMT45444016,PRE69056216,JPAMT42153001,
PRL116133903}, and skin phenomenon~\cite{PRL121026808,PRL121136802,PRL121086803,JPC2035043}.

Non-Hermiticity may bring new contents for carrying out the topological classification of matter and proper 
topological invariants may be defined in a new way~\cite{PRX8031079,PRB98115135,PRL121086803,
PRA97052115}. For one-dimensional topological systems with chiral symmetry, winding numbers are usually 
employed to characterize the topological properties~\cite{PRL622747,PRA98052116,EPL11210004}. 
Besides topological invariants defined in term of eigenstates, one may conjecture that complex eigenenergy 
spectra~\cite{PRL118040401,PRL120146402} and expectation values of operators may reveal topological 
features of matter. 
 
Non-Hermitian degenerate points with coalesced eigenenergies and eigenvectors may manifest themselves 
as exceptional points, lines, and surfaces. Non-Hermitian systems usually perform amusingly and extraordinarily 
at such points and nearby. For example, unidirectional invisibility has been observed in parity-time symmetric 
fiber networks near exceptional points~\cite{Nature488167}. Exceptional lines in nodal-line semimetals have 
been investigated in~\cite{PRB99081102,PRB99075130}. Weyl exceptional rings can occur in three-dimensional 
dissipative cold atomic gas~\cite{PRL118045701}. Non-Hermitian nodal semimetals are promoted to be 
symmetry-protected, where the surfaces of exceptional points form the boundaries of open Fermi volumes
\cite{PRB99041406}. Besides, the exceptional surface has been studied by symmetry-preserving non-Hermitian 
deformations of topological nodal line~\cite{arxiv181006549}. Intriguing properties make exceptional points 
valuable to be investigated in this work.

The bulk-boundary correspondence bridges the appearance of the edge states to the bulk topological invariants. 
It is intriguing to examine this correspondence in the non-Hermitian realm. Topological invariants characterizing 
the anomalous helical edge states of non-Hermitian Chern insulator has been discussed in~\cite{PRB98165148}. 
With the breakdown of conventional bulk-boundary correspondence, non-Bloch Chern numbers predicting the 
numbers of chiral edge modes were introduced in~\cite{PRL121136802}. Based on the notion of biorthogonal 
quantum mechanics, generalized bulk-boundary correspondence were studied for the non-Hermitian SSH model 
and Chern insulators~\cite{PRL121026808}.

There are currently platforms which can be employed to investigate non-Hermitian topological properties. For 
example, non-Hermitian topological phase transition has been investigated in the system of coupled optical 
waveguides~\cite{PRL115040402}. Photonic crystal systems also offer controllable platforms to investigate 
non-Hermitian topological properties of matter~\cite{NM16433,Nature565622}. 

The immunity against imperfections makes topological edge states appropriate for topological quantum 
computation. Topological stability of such states has been investigated for non-Hermitian systems
\cite{PRB84205128,PRL102065703}. Topological phase and the associated protection phenomena can 
emerge in the quantum wire of spinless fermions in an optical lattice in the presence of engineered dissipation
\cite{NP7971}.

In this work, we consider a crosslinked chain to study the topological features in the presence of balanced 
on-site gain and loss. The topological phase diagram would be given in term of a winding number versus 
system parameters. And the relation between this phase transition and appearance of exceptional points 
would be examined. Topological edge states locating mainly at the joints between domains of different 
topological phases hold on with elongating of the chain and conform to the bulk-boundary correspondence 
in this case. With strengthening of the balanced gain and loss, the chain will be pushed into the topologically 
trivial phase from the nontrivial phase. Such a topological feature can also be reflected by the vortex 
structures in the vector fields of complex eigenenergies or expectations of Pauli operators and their trajectories. 
The robustness of the edge states against various disorders would be discussed in terms of the experimental 
implementations for this model.

This work is organized as follows, in Sec.\ref{S:Model}, we put forward the crosslinked chain in the 
presence of on-site balanced gain and loss. In Sec.\ref{S:TPD}, the phase diagram in term of a winding 
number versus parameters is given with exceptional points, edge states and bulk-boundary correspondence 
discussed. In Sec.\ref{S:NHTPT}, we show the influence of non-Hermiticity on the energy spectra. Complex 
eigenenergies and expectation values of Pauli matrices are employed to reflect the topological feature. In 
Sec.\ref{S:Disorder}, the experimental implementations are proposed and the immunity of the edge states 
against several disorders is examined.  At last, we conclude in Sec.\ref{S:Concl}.
\begin{figure}[t]
\vspace{-1em}
\centering
\includegraphics[width=0.5\textwidth]{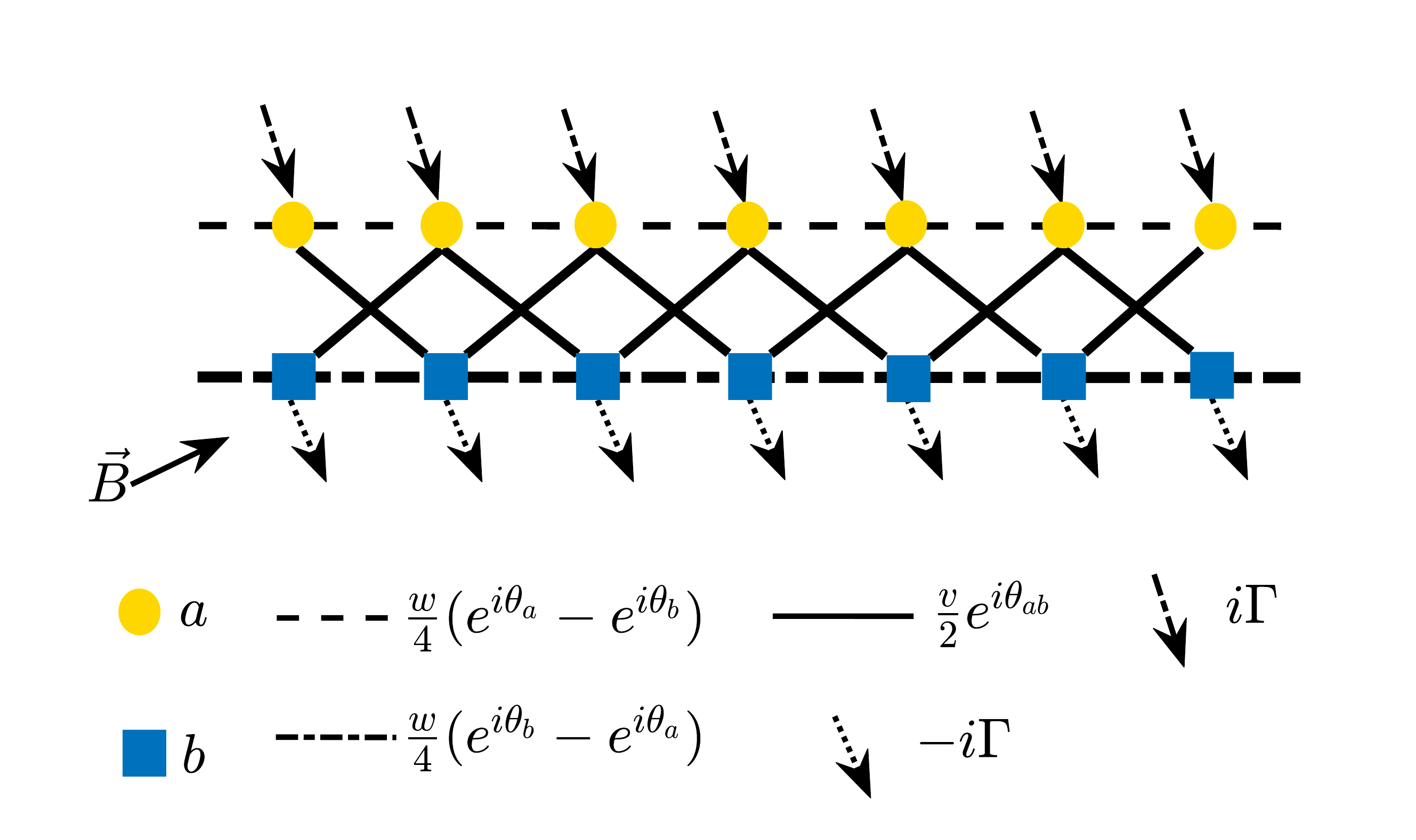}
\caption{The sketch of the crosslinked chain. $\textbf{a}$ sites are the gain sites with potential $i\Gamma$ 
and $\textbf{b}$ denote the loss sites with potential $-i\Gamma$.}
\label{f:modelchain}
\end{figure}
\section{The non-Hermitian crosslinked chain}
\label{S:Model}
We consider the lattice model composed of two-component fermions $\textbf{a}$ and $\textbf{b}$ 
without considering the spin freedom. The Hamiltonian in real space reads
\begin{eqnarray}
\begin{split}
&H=\sum\limits_{n} \frac{w(e^{i\theta_a}-e^{i\theta_b})}{4}(a_{n}^{\dag}a_{n+1}
-b_{n}^{\dag}b_{n+1})\\
&~~~+\frac{v}{2}e^{i\theta_{ab}}(a^{\dag}_{n}b_{n+1}+b^{\dag}_{n}a_{n+1})+h.c.+H_{nH} ,\\
&H_{nH}=i\Gamma(a^{\dag}_{n}a_n-b^{\dag}_{n}b_n),
\label{HamiltonianR}
\end{split}
\end{eqnarray}
where $a_n$ and $b_n$ are the fermionic annihilation operators for $\textbf{a}$ and $\textbf{b}$ elements 
on cell $n$ as Figure~\ref{f:modelchain} shows. $w$ is the coupling strength and $v$ would be used as the 
energy unit in this work. While this model is regarded as a simpliafied Creutz ladder~\cite{PRL832636}, where 
for electrons on a chain, the hopping potentials result from integral of gauge potential along a path connecting 
adjacent sites~\cite{PRL832636}. Since the magnetic field $\vec{B}$=$\nabla\times(\vec{A}+\nabla\Phi)$ with 
$\nabla\Phi$ freely added, the freedom of gauge choice makes the phases tunable. Without considering the 
spin freedom, such a model may be simulated by coupled arrays of waveguides or photonic crystals~\cite{JOSA55261,OE2006146055,NP7907,NN12675,PRL115040402,NM16433,Nature565622} which will be 
discussed at last.

Under periodic boundary condition, through the Fourier transformation $c_n=\frac{1}{N}\sum_k c_ke^{ikn}$ 
($c_{n(k)}$ represents the operators $a_{n(k)}$ or $b_{n(k)}$), in momentum space, the Hamiltonian can 
be written as
\begin{eqnarray}
H_{k} =h_z\sigma_z+h_x\sigma_x,
\label{eq:Hk}
\end{eqnarray}
where $\sigma_z$ and $\sigma_x$ are Pauli matrices and $h_z$=
$w\sin(k+\frac{\theta_{a}+\theta_{b}}{2})\sin(\frac{\theta_{b}-\theta_{a}}{2})$+ 
$i\Gamma$, $h_x$=$\cos(k+\theta_{ab})$. 

 Topological properties are usually related to eigenstates of systems~\cite{PRL622747}. The right eigenstates 
 of $H_k$ can be calculated by solving $H_{k}|u_{R,i}\rangle$=$E_k|u_{R,i}\rangle$ ($i$=1,2)~\cite{MC}. The 
 left eigenstates $\langle u_{L,i}|$ can be obtained by calculating the right eigenstates of $H_k^{\dagger}$. And 
 the energy dispersion  relation can be obtained simultaneously. In the Hermitian situation, the degenerate points 
 in the energy  dispersion are closely related to a topological phase transition~\cite{RMP823045,RMP831057}. 
 Analogously, one may conjecture that similar relation may hold on in the non-Hermitian realm. Merely the 
 degenerate points  are referred as exceptional points with a defective non-Hermitian Hamiltonian. In this case, 
 the non-Hermiticity originates from the on-site gain and loss. We mainly focus on the influence of such non-Hermitian 
 terms on the topological properties of this model.
\begin{figure}[htb!]
	\includegraphics[width=0.5\textwidth]{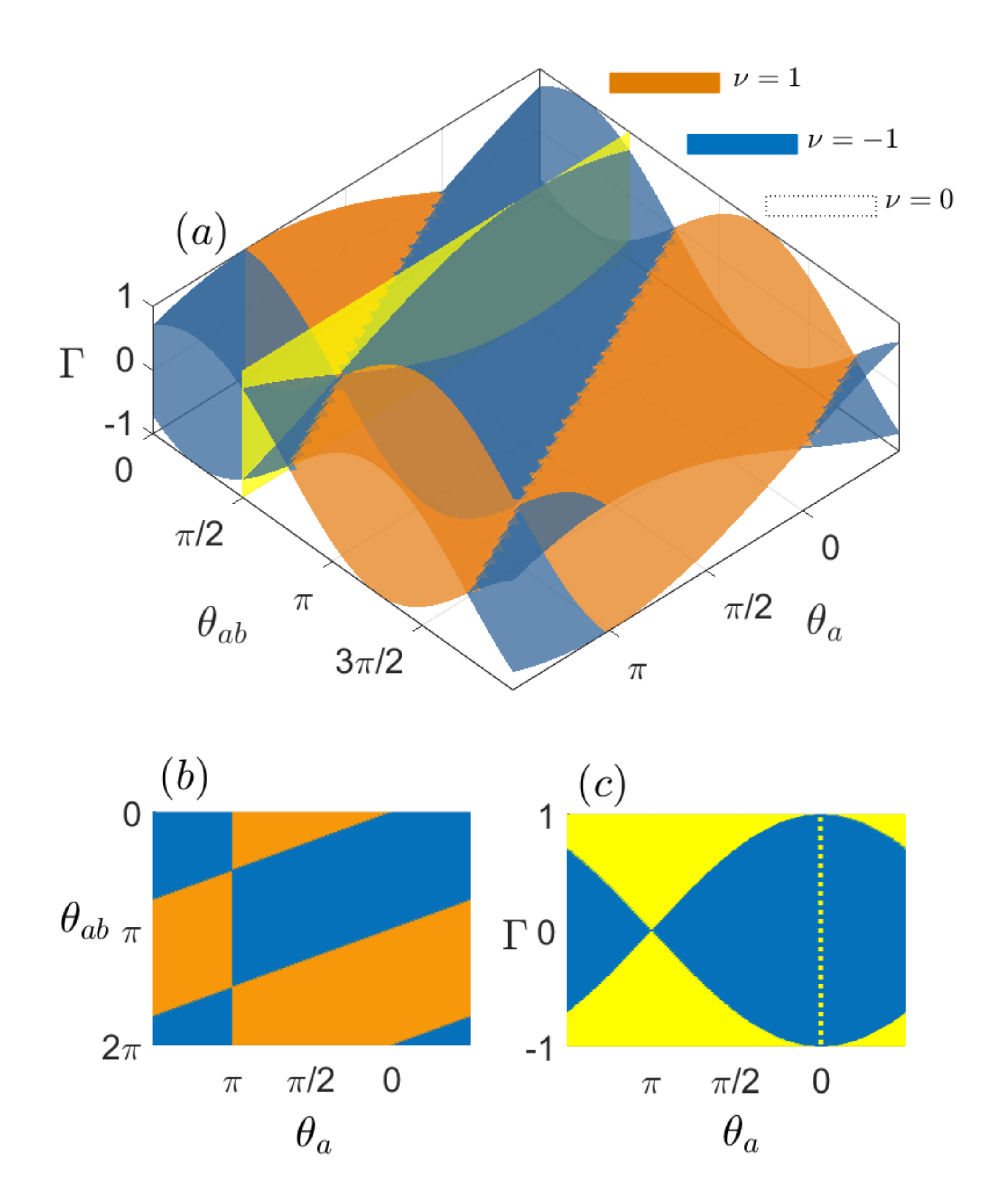}
	\vspace{-0.5cm}
	\caption{$(a)$ Phase diagram in term of  $\nu$ defined in (\ref{vstotal}) as a function of $\theta_a$, 
	$\theta_{ab}$ and $\Gamma$ when $\theta_{b}=\pi$. The value of the winding number is mapped to the 
	colors as shown in $(a)$. $(b)$ Phase diagram as a function of $\theta_a$ and $\theta_{ab}$ when 
	$\Gamma$=0. $(c)$ Phase diagram as a function of $\theta_a$ and $\Gamma$ when $\theta_{ab}$=
	$\pi/2$ as the yellow plane shows in $(a)$.}
	\label{f:PTG}
\end{figure}

\section{Topological Phase Diagram and Edge States}
\label{S:TPD}
\subsection{Topological Phase Diagram}
\label{Ss:TPD}
Firstly, we check the topological phase diagram of this non-Hermitian crosslinked chain in term of a winding 
number versus hopping phases and strength of the balanced gain and loss in the Hamiltonian (\ref{eq:Hk}). 
Besides the relation between topological phase transition and appearance of exceptional points, 
bulk-boundary correspondence correlating topological invariants to edge states under open boundary 
condition is also intriguing to be examined especially in the presence of gain and loss in this model. 
\begin{figure*}[htb!]
	\includegraphics[width=1\textwidth]{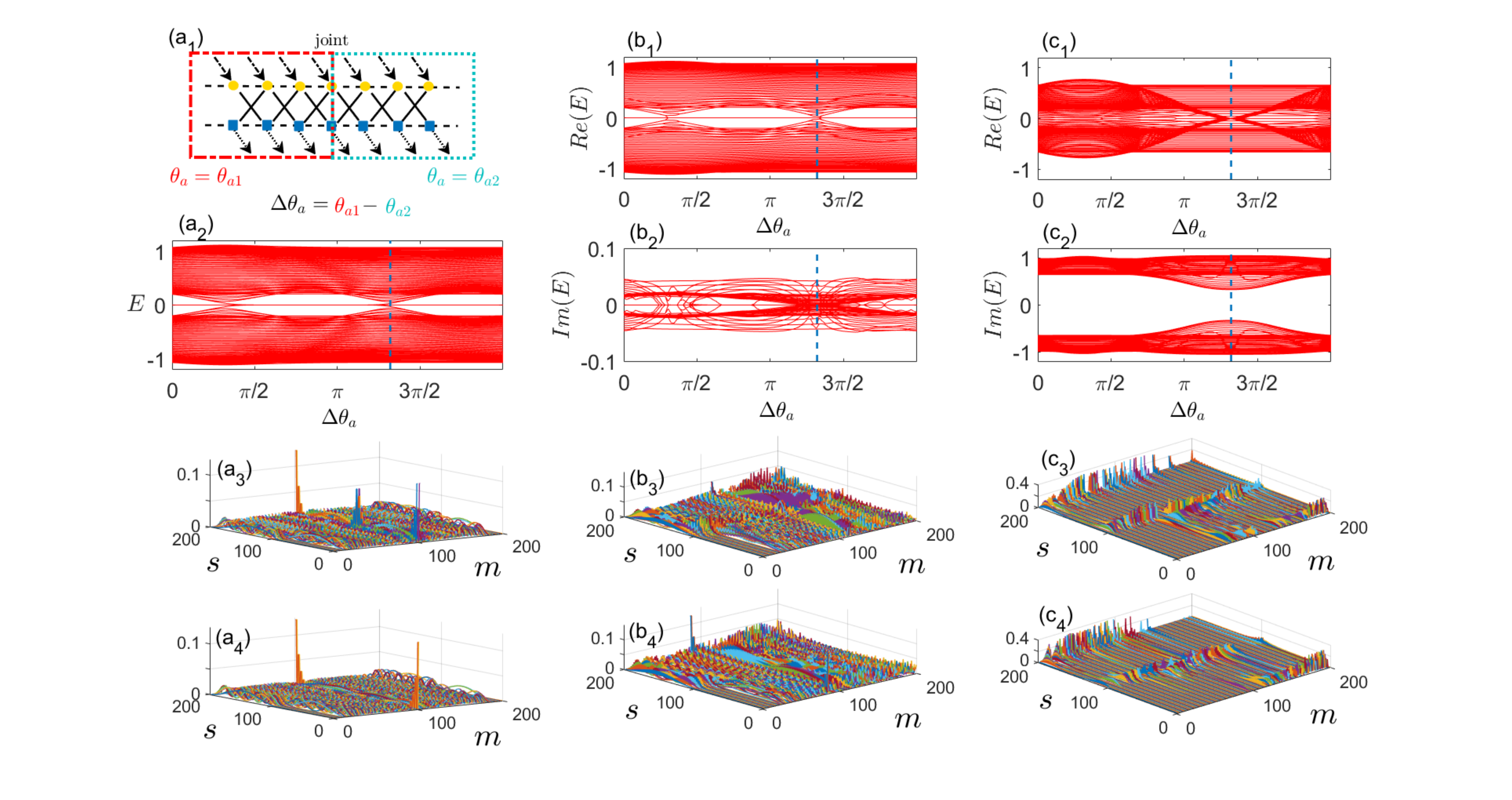}
	\vspace{-1cm}
	\caption{$(a_1)$ Sketch of the chain jointed by two chains with different $\theta_a$ ($\theta_{a1}$ and 
	$\theta_{a2}$, respectively). $(a_2)$ Energy spectrum versus the phase difference: $\Delta\theta_{a}$
	=$\theta_{a1}-\theta_{a2}$ under open boundary condition when $\Gamma=0$ and $\theta_{a2}$ fixed. 
	$(a_3)$ Distributions of the eigenstates (denoted by $m$) on the chain sites (denoted by $s$) when 
	$\Delta\theta_{a}=3.5$ ($\nu$=1 for the chain with $\theta_{a1}$). $(a_4)$ Distributions like $(a_3)$ except 
	$\Delta\theta_{a}=5$ ($\nu$=-1 for the chain with $\theta_{a1}$). The other parameters are $\theta_{a2}$
	=-1, $\theta_b=\pi$, $\theta_{ab}=0.5$ and $w=0.5$ ($\nu$=-1). $(b_1)$ and $(b_2)$ The real and imaginary 
	eigenenergy spectra when $\Gamma=0.05$. $(b_3)$ and 	$(b_4)$ The distributions of the eigenstates with 
	the same parameters and $\nu$ to $(a_3)$ and $(a_4)$ except $\Gamma=0.05$.	$(c_1)$ - $(c_4)$ are same 
	to $(b_1)$ - $(b_4)$ except $\Gamma=1.05$ ($\nu=0$). The vertical dashed lines show the value 
	$\Delta\theta_a$=$\pi$-$\theta_{a2}$, where the topological phase transition occurs according to \ref{f:PTG}.}
	\label{f:CLIPR6} 
\end{figure*}

Considering the Hamiltonian (\ref{eq:Hk}), the winding number is defined as 
\begin{eqnarray}
\begin{split}
\nu=\frac{1}{2\pi}\oint_{BZ}\frac{h_z\partial_{k}h_x-h_x\partial_{k}h_z}{h_x^2+h_z^2} dk =
\frac{1}{2\pi}\oint_{BZ} \partial_k\phi dk\\
\label{vstotal}
\end{split}
\end{eqnarray} 
in the Brillouin zone (subscript $BZ$) where $\phi$=$atan\frac{h_x}{h_z}$. Thus $\nu$ has the meaning 
of accumulation of the angle $\phi$ as $k$ sweeps the Brillouin zone. The complex angle $\phi$=
$Re(\phi)+i Im(\phi)$=$\phi_r+ i \phi_i$. Then $\nu$=$\frac{1}{2\pi}\oint \partial_k\phi_r dk$+ 
$i \frac{1}{2\pi}\oint \partial_k\phi_idk$. Since $e^{i2\phi}$= $\frac{h_z+i h_x}{hz-i h_x}$, it can be 
calculated that 
 \begin{eqnarray}
 \begin{split}
 &\phi_r=\frac{1}{2}atan\frac{2(h_{xr}h_{zr}+h_{xi}h_{zi})}{|h_z|^2-|h_x|^2},\\
 \\
 &\phi_i=-\frac{1}{4}log\sqrt{\frac{|h_x|^2+|h_z|^2-2(h_{xi}h_{zr}-h_{xr}h_{zi})}{|h_x|^2+|h_z|^2+2(h_{xi}h_{zr}-h_{xr}h_{zi})}},\\ 
\label{phir}
\end{split}
\end{eqnarray} 
where ($h_{\alpha,r(i)}$ denotes real (imaginary) parts of $h_{\alpha}$ and $\alpha$=$x$,$z$, $i$=1,2). After 
some calculations similar to~\cite{SB631385}, we obtain $\frac{1}{2\pi}\oint_{BZ} \partial_k\phi_idk=0$ and 
 \begin{eqnarray}
 \begin{split}
 \nu&=\frac{1}{2\pi}\oint_{BZ} \partial_k\phi_r dk\\
     &=\frac{1}{2}sgn(\cos(\theta_{ab}-\frac{\theta_a+\theta_b}{2})\sin(\frac{\theta_b-\theta_a}{2}))\\
     &(sgn(\Gamma^2-\cos^2(\theta_{ab}-\frac{\theta_a+\theta_b}{2}))-\eta),
 \end{split}
 \label{nutotal}
\end{eqnarray} 
where $\eta$=$sgn(\Gamma^2+w^2\cos^2(\theta_{ab}-\frac{\theta_a+\theta_b}{2})\sin^2(\frac{\theta_b-\theta_a}{2}))$
$\geq0$ and $sgn(x)$ outputs the sign of $x$. As shown in Figure~\ref{f:PTG}, this model is topological nontrivial 
for the full range of $\theta_a$ and $\theta_{ab}$ when $\theta_b=\pi$ and $\Gamma=0$.

In the topological classification of matter, degeneration in energy dispersion usually accompanies with a topological 
phase transition. According to $\nu$ in (\ref{nutotal}), the phase boundaries occur at $\Gamma^2$=
$\cos^2(\theta_{ab}-\frac{\theta_a+\theta_b}{2})$ and $(\theta_b-\theta_a)$=$2m\pi$ ($m$ is an integer). This is 
supported in the Hermitian case as shown in Figure \ref{f:PTG}(b), namely, the phase boundaries versus the 
parameters are consistent with the trajectories of degenerate points of the Hamiltonian (\ref{eq:Hk}) when 
$\Gamma$=0. Similarly, we compare this phase transition and the appearance of exceptional points in the 
non-Hermitian case. For this two-band model when $H_k\neq0$, exceptional point occurs when $H_k$ is defective, 
namely, the eigenenergies and eigenstates both coalesce with orthogonality between the left and right eigenstates. 
Thus, after some algebra, one can find that the exceptional points locate on the surfaces 
$\Gamma^2$=$\cos^2(\theta_{ab}-\frac{\theta_a+\theta_b}{2})$ and $\theta_b-\theta_a=2m\pi$ ($m$ is an 
integer). These exceptional surfaces coincide with the phase boundaries of $\nu$ mentioned above. Exceptional 
points on the surfaces $\Gamma^2$=$\cos^2(\theta_{ab}-\frac{\theta_a+\theta_b}{2})$ fulfill the relation 
$k=m\pi-\frac{\theta_a+\theta_b}{2}$ and $\Gamma^2=\cos^2(k+\theta_{ab})$. Those on the surfaces 
$\theta_b-\theta_a$=$2m\pi$ fulfill $\Gamma^2$=$\cos^2(k+\theta_{ab})$ as shown in Figure~\ref{f:PTG}(a) 
when $\theta_a=\pi$. In both cases, the Hamiltonian $H_k$ turns to be proportional to the defective matrix 
$i\sigma_z\pm\sigma_x$. The exceptional points also separate topological phases in this case, similar to the 
degenerate points in the Hermitian realm. 

\subsection{Topological Edge States}
\label{Ss:TES}
Topological edge states with distributions mainly near the joints between domains with different topological 
invariants comply with the bulk-boundary correspondence. And one may conjecture that such edge states 
resulting from topology hold on with elongating the chain. In view of the winding number $\nu$, the vacuum 
can be regarded as a topological trivial domain which results to edge states for topological nontrivial systems 
under open boundary condition. Jointing chains with the same topological invariant but different parameters, 
there should be no edge states with distribution mainly at the joints on the long chain. To check this conjecture 
in non-Hermitian realm, we joint two chains with different $\theta_a$ and $\Gamma$ into one chain under 
open boundary condition as shown in Figure~\ref{f:CLIPR6}$(a_1)$. 

The energy spectra as a function of the difference between the two $\theta_a$s in the two parts are shown 
in Figure~\ref{f:CLIPR6} ($a_2$), ($b_{1(2)}$) and ($c_{1(2)}$) for different $\Gamma$s. The Energy 
spectra touch at the phase transition points ($\theta_{a2}$ remains constant in Figure~\ref{f:CLIPR6}) in 
the gap which coincides with Figure~\ref{f:PTG}. And the symmetry of the spectra versus zero energy results 
from the symmetry of this model: $\sigma_y H_k\sigma_y$=$-H_k$. So if $H_k$ has an eigenvector $|u\rangle$ 
with eigenvalue $E$, then $\sigma_y|u\rangle$ is also an eigenvector with eigenvalue $-E$. In real space, this 
symmetry reads $\bigoplus_n \sigma_{y,n}$ applied to the counterpart of the spinless Hamiltonian obtained by 
Jordan-Wigner transformation~\cite{Sachdev}.

As shown in Figure \ref{f:CLIPR6} ($a_{3(4)}$), ($b_{3(4)}$) and ($c_{3(4)}$), it can be found that no edge 
states exist at the joints between domains with different $\theta_a$ but same $\nu$ when the chain is very long 
in the topological nontrivial realm. Yet we numerically find that the localized states may appear with distributions 
mainly around the joints when the chain is short. But such states would submerge into the eigenstates with 
elongating the chain. On the other hand, around the joints between domains with different $\nu$, the edge states 
hold on obviously even the chain is very long which confirms their topological origin. 

In the non-Hermitian case, with increasing of $\Gamma$, the eigenstates tend to distribute locally on fractional 
sites instead of dispersedly on the chain. When $\Gamma$ pushes the chain to the topologically trivial phase, 
the eigenstates perform skin effect~\cite{PRL121026808,PRL121136802,PRL121086803,JPC2035043}, namely, 
most of the states tend to distribute at the joints between domains with different parameters and the edge states 
would submerge into these localized states. 
\begin{figure}[htb!]
	\includegraphics[width=0.5\textwidth]{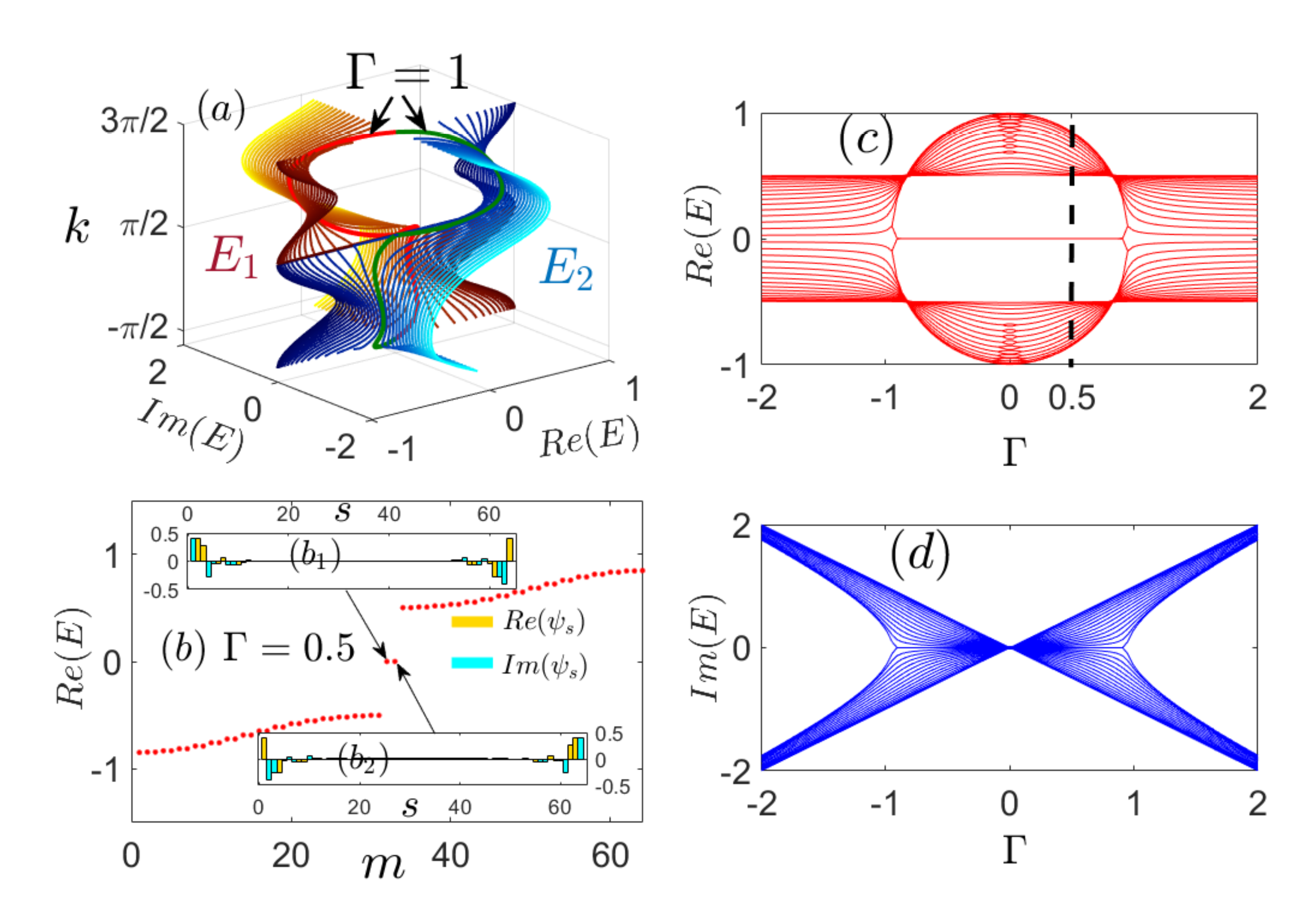}
	\vspace{0cm}
	\caption{$(a)$ The energy spectra for $\Gamma\in[0,2]$ (darker color corresponds to less $\Gamma$) in 
	momentum space when $(\theta_a,\theta_b,\theta_{ab},w)$=$(0, \pi, 0.5\pi,0.5)$. $(b)$ The real part of 
	eigenenergies as the dashed vertical line in (c) shows. The insets $(b_1)$ and $(b_2)$ provide the profiles 
	of the real and imaginary parts of the edge states $|\psi_{s}\rangle$ projecting on the chain. $(c)$ and 
	$(d)$ The real and imaginary parts of the 	eigenenergy spectrum under open boundary condition.}
	\label{f:CLExcep} 
\end{figure}
\section{Topological Features in Complex Energies and Expectation values of Operators}
\label{S:NHTPT}
\subsection{Complex Spectra}
\label{ESGL}
To get insight into the influence of the balanced gain and loss on the topological feature more concretely, 
we focus on the chain with parameters $(\theta_a,\theta_b,\theta_{ab})$=$(0, \pi, 0.5\pi)$ as the dashed 
line shows in Figure~\ref{f:PTG} (c) with the phase transition occurring at $\Gamma=\pm1$. It may benefit 
revealing the topological phase transition by checking the energy spectra versus $\Gamma$ both in 
momentum and real spaces. Such complex eigenenergy spectra versus $\Gamma$ are shown in Figure
\ref{f:CLExcep}. $(\Gamma,k)=(\pm1,\frac{2n+1}{2}\pi)$ ($n$ an integer) are the points where the 
trajectories of the complex energies cross. Such points in momentum space are exceptional points with 
coalesced eigenstates and eigenenergies of the defective Hamiltonian.  

Bulk-boundary correspondence is an intriguing issue when exploring topological properties for Hermitian 
systems~\cite{RMP823045,RMP831057}. Non-Hermitian terms in Hamiltonians usually raise complex energy 
spectrum which makes the problem intricate. In this model, the results in Figure~\ref{f:CLExcep} conform 
to the bulk-boundary correspondence. In Figure~\ref{f:CLExcep}(b), we show the real eigenenergy as the 
dashed vertical line shows in Figure~\ref{f:CLExcep}(c). The insets $(b_1)$ and $(b_2)$ show the real 
and imaginary parts of amplitudes of the edge states projecting on the chain sites. The real (imaginary) part 
of the amplitude of one edge state is identical to the imaginary (real) part of the other one.
\begin{figure}[htb!]
\vspace{-0.5cm}
\includegraphics[width=0.5\textwidth]{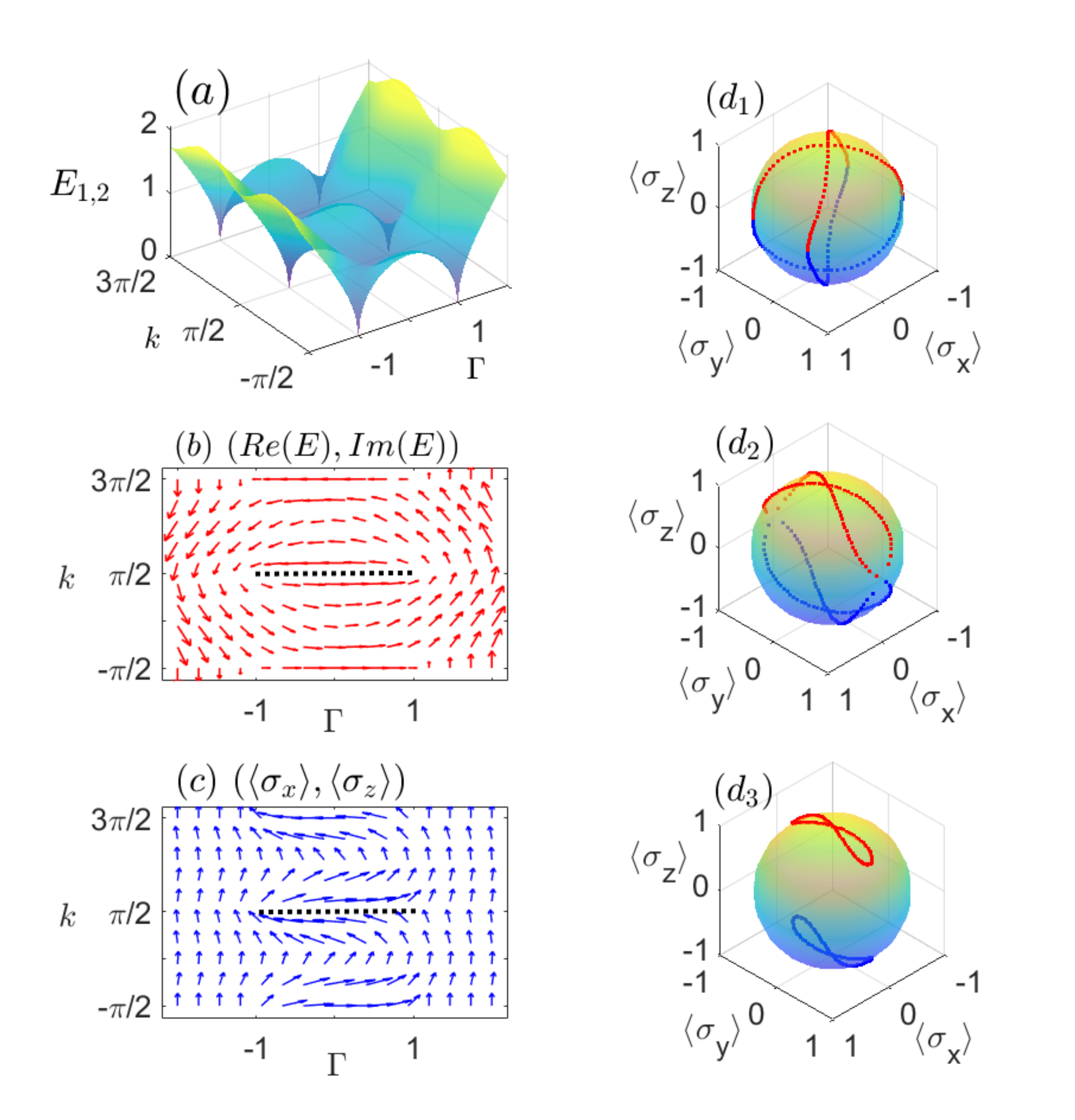}
\caption{$(a)$ The absolute value of $E_{1}$ (or $E_{2}$) versus $k$ and $\Gamma$. $(b)$ and $(c)$ 
The vector fields of $(Re(E),Im(E))$ and $(\langle\sigma_x\rangle,\langle\sigma_z\rangle)$ on the grid of 
$(\Gamma,k)$. $(d_1)-(d_3)$ The trajectories of the points with coordinates 
$(\langle\sigma_x\rangle,\langle\sigma_y\rangle,\langle\sigma_z\rangle)$ for $k\in[0,2\pi)$ when 
$\Gamma$=(0.5, 1, 1.5), respectively. The other parameters are $(\theta_a, \theta_b, \theta_{ab})$=
$(0, \pi, 0.5\pi)$ as the vertical dashed line shows in Figure~\ref{f:PTG} (c).}
\label{f:nonExpected}
\end{figure}
\subsection{Topological feature reflected by vortex}
\label{s:Vortex}
Topological classification of matter in terms of topological invariants is usually defined by integrals in a global 
manner~\cite{RMP823045,RMP831057}. Besides, for non-Hermitian systems, complex energy 
may result to specific vector structures reflecting the topological properties, merely ambiguity of band-labeling 
occurs at the exceptional points. We set the eigenenergies with the same sign of imaginary part versus $k$ 
in the Brillouin zone as in one branch as in Figure~\ref{f:CLExcep} (a). Since a complex number $x$ can be 
visually represented by a pair of real numbers ($Re(x)$,$Im(x)$), we employ such pairs of numbers to construct 
a vector field for the complex eigenenergies represented by arrows on the parameter grid as in Figure
\ref{f:nonExpected} (b). The vector field of complex energy $(Re(E),Im(E))$ has half-vortex structures at the 
exceptional points $(\Gamma,k)=(\pm1,\frac{2n+1}{2}\pi)$ with the branch cut between $(-1,\frac{2n+1}{2}\pi)$ 
and $(1,\frac{2n+1}{2}\pi)$ on the grid of $(\Gamma,k)$. It should be emphasized that the structures 
of the half-vortexes and the position of branch cut are related to the choice of energy branches due to the 
ambiguity of band-labeling at the exceptional points~\cite{PRL118040401}. For example, while one sets the 
eigenenergies with the same sign of real part (opposite sign of imaginary part) versus $k$ as in one branch, 
the location of the center of the half-vortexes do not change but the branch cut would change to 
$(-\infty,\frac{2m+1}{2}\pi)$-$(-1,\frac{2n+1}{2}\pi)$ and $(1,\frac{2n+1}{2}\pi)$-$(+\infty,\frac{2n+1}{2}\pi)$ on 
the grid of $(\Gamma,k)$. 

Similarly, half-vortex structures also appear in the vector field of ($\langle\sigma_x\rangle$,$\langle\sigma_z\rangle$) 
at the exceptional points where $\langle\sigma_{\xi}\rangle$=$\langle u_{L,i}|\sigma_{\xi}|u_{R,i}\rangle$ ($\xi$= $x$, 
$y$ or $z$ and $i$=$1$ or $2$) as shown in Figure~\ref{f:nonExpected} (c).
\subsection{Topological feature reflected by trajectory}
\label{s:Trajectory}
Considering the eigenenergies are complex values, we conjecture that a topological object may be identified 
from the trajectory of the complex spectrum, with no counterparts in Hermitian realm. In the supplementary 
material `CrossLinkavi.avi', we show the loops of $(\Gamma,k)$ and the corresponding trajectories of 
$(Re(E_{1,2}),Im(E_{1,2}))$. We can see that the loops of the complex eigenenergies do not connect into 
one loop unless the loop of $(\Gamma,k)$ encircles one of the exceptional points $(\Gamma,k)=(1,\frac{\pi}{2})$ 
independent to the shape of the $(\Gamma,k)$. Otherwise, the loops of the eigenenergies separate into two 
loops. As mentioned in Section (\ref{S:TPD}), exceptional points play the role of topological critical points. This 
is similar but different to characterize topological phase transition by defining a topological invariant on a closed 
curve surrounding the phase transition point in the parameter space in~\cite{PRB92085118}. 

Similarly, we find that the trajectories of the expectations of Pauli matrices on the two eigenstates as 
$(\langle\sigma_x\rangle,\langle\sigma_y\rangle,\langle\sigma_z\rangle)$ in one Brillouin zone can reflect 
the topological feature of this model as shown in Figure~\ref{f:nonExpected} ($d_1$)-($d_3$). When 
$|\Gamma|<1$,  the two trajectories connect to each other with two cross points on the sphere. But when 
$|\Gamma|>1$, the trajectories divided into two non-connected loops with two crosses. $|\Gamma|=1$ are 
the topological critical points with four crosses on the sphere. This is consistent with the phase diagram 
in Section (\ref{S:TPD}) and the energy spectra in Figure~\ref{f:CLExcep}.

Although the general constraint conditions for the operators used to reflect topological features for a topological
model need further investigation, these results hint that besides winding number $\nu$, other quantities can 
play the role to reflect topological properties of matter. This may shed light on revealing the topological phase 
of matter by accessible manner like mentioned above.
 \begin{figure}[t]
 	\includegraphics[width=0.5\textwidth]{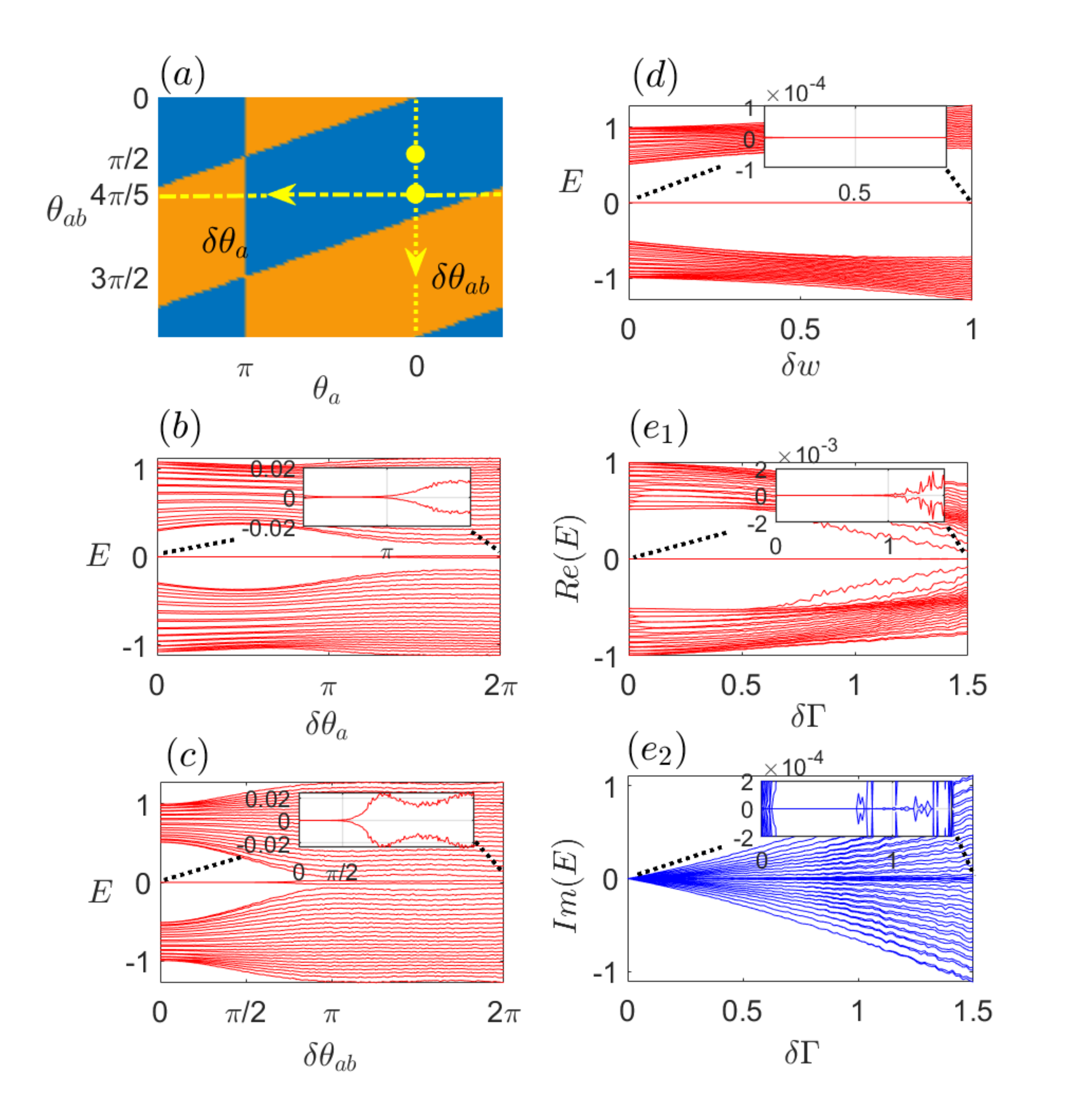}
 	\vspace{-0.5cm}
 	\caption{$(a)$ The variation of the amplitudes $\delta\theta_a$ and $\delta\theta_{ab}$ in the phase diagram 
 	when $\Gamma$=0. The horizontal arrow shows the variation of $\delta\theta_a$ and the vertical one shows 
 	that for $\delta\theta_{ab}$ with the big dots indicate the initial parameters. $(b)$, $(c)$ and $(d)$ The energy 
 	spectrum versus $\delta\theta_a$, $\delta\theta_{ab}$ and $\delta w$, respectively for the chain when 
 	$\Gamma$=0 under open boundary condition. $(e_1)$-$(e_2)$ The real and imaginary parts of energy 
 	spectrum versus $\delta \Gamma$. The initial parameters for $d$, $e_1$ and $e_2$ are same as ($\theta_a$, 
 	$\theta_b$, $\theta_{ab}$, $\Gamma$, $w$)=(0, $\pi$, 0.5$\pi$, 0, 0.5). Each panel is averaged over 100 
 	independent 
   simulations.}
 	 \label{f:disorder}
 \end{figure}
 \section{Experimental implementations and robustness of edge states against disorders}
\label{S:Disorder}
Experimentally, coupled optical waveguides written in bulk materials by laser written techniques\cite{JOSA55261,OE2006146055,NP7907,NN12675,PRL115040402} or photonic crystals~\cite{NM16433,
Nature565622} may be candidates to fabricate the simulators of this chain. In these proposals, the coupling 
between sites is the essential factor. The hopping strengths and phases can be tuned by varying the coupling 
conditions between the sites. The gain can be obtained by coupling the simulators to gain medium or pumping. 
And the loss can be introduced naturally by scattering due to impurity or defects in the materials. 

In terms of the real implementations mentioned above, various kinds of disorders are usually inevitable due to 
defects or impurities in the materials. The robustness of topological edge modes against such disorders is an 
advantage in the view of applications. Such states usually manifest themselves as gaped states as the energy 
bands shows in this model in Figure~\ref{f:CLExcep}. The edge states should not blend into the bands until 
the disorders exceed a threshold. Considering the Hamiltonian and the implementations of this model mentioned 
above, we mainly focus on the disorders of $\theta_a$, $\theta_{ab}$, $w$ and $\Gamma$. The simulation 
results are shown in Figure~\ref{f:disorder}. In each simulation, the disorders distribute on all the cells of the 
chain with strength randomly and uniformly within the range $[0,\delta x]$ ($\delta x$ represents the maximal 
amplitudes for the disorders in $\theta_a$, $\theta_{ab}$, $w$ and $\Gamma$). The parameters in the 
Hamiltonian (\ref{HamiltonianR}) become $x+\delta x$ in each simulation. It can be seen that the edge states 
behave robustly to the disorders until the phase transition occurs, namely, passing the exceptional points. Since 
the amplitude of $w$ does not change $\nu$ in (\ref{nutotal}), the edge states hold on with increasing of $\delta w$.
It should be noted that the amplitudes of the disorders should not be too large since they are perturbations here. 
\section{Conclusion}
\label{S:Concl}
Topological phase transition of a crosslinked chain in the presence of balanced gain and loss has been 
investigated in this work. The phase transition is indicated by a winding number and the phase boundaries 
coincide with the surfaces of exceptional points. The edge states with distributions mainly near the joints 
between domains of different phases conform to the bulk-boundary correspondence and the skin effect is 
also found with increasing of the balanced gain and loss. Besides the winding number, the topological feature 
of this model not only can be reflected by the vortex structures in the vector fields of complex eigenenergies 
and expectation values of operators but also their trajectories. This model may be demonstrated by coupled 
waveguides or photonic crystals. And the edge states behave tenaciously against various disorders until the 
topological phase transition occurs. This work benefits our insight into the field of topological classification of 
matter in non-Hermitian realm and extends the avenue to reflect topological features of matter.
\section*{Acknowledgements}
The authors thank Shu Chen for helpful discussions. This work is supported by the National Natural Science 
Foundation of China (Grant No. 11725417, 11575027), NSAF (Grant No. U1730449), and Science Challenge 
Project (Grant No. TZ2018005).


\begin{thebibliography}{99}
\bibitem{RMP823045} M. Z. Hasan and C. L. Kane, Colloquium: Topologicalinsulators, Rev. Mod. Phys. 
\textbf{82}, 3045 (2010).

\bibitem{RMP831057} X. L. Qi and S. C. Zhang, Topological insulators and superconductors, Rev. Mod. 
Phys. \textbf{83}, 1057 (2011). 

\bibitem{PRL702273} H. J. Carmichael, Quantum trajectory theory for cascaded open systems, Phys. 
Rev. Lett. \textbf{70}, 2273 (1993).

\bibitem{PRL121213902} K. Takata and M. Notomi, Photonic Topological Insulating Phase Induced 
Solely by Gain and Loss,  Phys. Rev. Lett. \textbf{121}, 213902 (2018).

\bibitem{NP1411} R. Elganainy, K. G. Makris, M. Khajavikhan, Z. H. Musslimani, S. Rotter, and D. N. 
Christodoulides, Non-Hermitian physics and PT symmetry, Nat. Phys. \textbf{14}, 11 (2018).

\bibitem{NC10297} K. Kawabata, S. Higashikawa, Z. Gong, Y. Ashida, and M. Ueda, Topological 
unification of time-reversal and particle-hole symmetries in non-Hermitian physics, Nat. Commun. \textbf{10}, 
297 (2019).

\bibitem{PRE61929} W. D. Heiss, Repulsion of resonance states and exceptional points, Phys. Rev. E
\textbf{61}, 929 (2000).

\bibitem{CJP541039} M. V. Berry, Physics of non-Hermitian degeneracies, Czech. J. Phys. \textbf{54},
1039 (2004).

\bibitem{PRL86787} C. Dembowski, H.-D. Gr\"af, H. L. Harney, A. Heine, W. D. Heiss, H. Rehfeld, and 
A. Richter, Experimental observation of the topological structure of exceptional points, Phys. Rev. Lett. 
\textbf{86}, 787 (2001).

\bibitem{PNAS1136845} B. Peng, S. K. \"Ozdemir, M. Liertzer, W. Chen, J. Kramer, H. Yilmaz, J. Wiersig, 
S. Rotter, and L. Yang, Chiral modes and directional lasing at exceptional points, PNAS \textbf{113}, 6845 (2016).

\bibitem{PRL104153601} Y. Choi, S. Kang, S. Lim, W. Kim, J.-R. Kim, J.-H. Lee, and K. An, Quasieigenstate 
Coalescence in an Atom-Cavity Quantum Composite, Phys. Rev. Lett. \textbf{104}, 153601 (2010).

\bibitem{Nature525354} B. Zhen, C. W. Hsu, Y. Igarashi, L. Lu, I. Kaminer, A. Pick, S.-L. Chua, J. D. Joannopoulos, 
and M. Soljacic, Spawning rings of exceptional points out of Dirac cones, Nature (London) \textbf{525}, 354 (2015). 

\bibitem{Nature526554} T. Gao, E. Estrecho, K. Y. Bliokh, T. C. H. Liew, M. D. Fraser, S. Brodbeck, M. Kamp, 
C. Schneider, S. H \"ofling, Y. Yamamoto, F. Nori, Y. S. Kivshar, A. G. Truscott, R. G. Dall and E. A. Ostrovskaya,
Observation of non-Hermitian degeneracies in a chaotic exciton–polariton billiard, Nature \textbf{526}, 554 (2015).

\bibitem{arxiv180809541} A. Cerjan, S. Huang, K. P. Chen, Y. Chong, and M. C. Rechtsman, Experimental 
realization of a Weyl exceptional ring, arXiv:1808.09541 (2018).

\bibitem{JPAMT45444016} W. D. Heiss, The physics of exceptional points, J. Phys. A Math. Theor. \textbf{45}, 
444016 (2012).

\bibitem{PRE69056216} C. Dembowski, B. Dietz, H.-D. Gr\"af, H. L. Harney, A. Heine, W. D. Heiss, and 
A. Richter, Encircling an exceptional point, Phys. Rev. E \textbf{69}, 056216 (2004).

\bibitem{JPAMT42153001} I. Rotter, A non-Hermitian Hamilton operator and the physics of open quantum 
systems, J. Phys. A Math. Theor. \textbf{42}, 153001 (2009).

\bibitem{PRL116133903} T. E. Lee, Anomalous Edge State in a Non-Hermitian Lattice, Phys. Rev. Lett. 
\textbf{116}, 133903 (2016). 

\bibitem{PRL121136802} S. Yao, F. Song and Z. Wang, Non-Hermitian Chern Bands,  Phys. Rev. Lett. 
\textbf{121}, 136802 (2018). 

\bibitem{PRL121026808} F. K. Kunst, E. Edvardsson, J. C. Budich, and E. J. Bergholtz, Biorthogonal 
Bulk-Boundary Correspondence in Non-Hermitian Systems, Phys. Rev. Lett. \textbf{121}, 026808 (2018).

\bibitem{JPC2035043} Y. Xiong, Why does bulk boundary correspondence fail in some non-hermitian 
topological models, Journal of Physics Communications \textbf{2}, 035043 (2018).

\bibitem{PRL121086803} S. Yao and Z. Wang, Edge States and Topological Invariants of Non-Hermitian 
Systems, Phys. Rev. Lett. \textbf{121}, 086803 (2018). 

\bibitem{PRX8031079} Z. Gong, Y. Ashida, K. Kawabata, K. Takasan, S.Higashikawa, and M. Ueda,
Topological Phases of Non-Hermitian Systems, Phys. Rev. X \textbf{8}, 031079 (2018).

\bibitem{PRB98115135} S. Lieu, Topological symmetry classes for non-Hermitian models and connections 
to the bosonic Bogoliubov–de Gennes equation, Phys. Rev. B \textbf{98}, 115135 (2018).

\bibitem{PRA97052115} C. Yin, H. Jiang, L. Li, R. L\"{u}, and S. Chen,  Geometrical meaning of winding 
number and its characterization of topological phases in one-dimensional chiral non-Hermitian systems,
Phys. Rev. A  \textbf{97}, 052115 (2018).

\bibitem{PRL622747} J. Zak, Berry's Phase for Energy Bands in Solids, Phys. Rev. Lett.  \textbf{62}, 
2747 (1989).

\bibitem{PRA98052116} H. Jiang, C. Yang, and S. Chen, Topological invariants and phase diagrams 
for one-dimensional two-band non-Hermitian systems without chiral symmetry, Phys. Rev. A \textbf{98}, 
052116 (2018).

\bibitem{EPL11210004} L. Li, C. Yang, and S. Chen, Hidden-symmetry–protected topological phases on a 
one-dimensional lattice,  Eur. Phys. Lett. \textbf{112}, 10004 (2015).

\bibitem{PRL118040401} D. Leykam, K.Y. Bliokh, C. Huang, Y. D. Chong, and F. Nori, Edge Modes, 
Degeneracies, and Topological Numbers in Non-Hermitian Systems, Phys. Rev. Lett. \textbf{118}, 040401 
(2017).

\bibitem{PRL120146402} H. Shen, B. Zhen, and L. Fu,  Topological Band Theory for Non-Hermitian 
Hamiltonians, Phys. Rev. Lett. \textbf{120}, 146402 (2018).

\bibitem{Nature488167}  A. Regensburger, C. Bersch, M. A. Miri, G. Onishchukov, D. N. Christodoulides, 
and U. Peschel, Parity–time synthetic photonic lattices,  Nature (London) \textbf{488}, 167 (2012).

\bibitem{PRB99081102} Z. Yang and J. Hu, Non-Hermitian Hopf-link exceptional line semimetals, Phys. Rev. 
B \textbf{99}, 081102 (2019).

\bibitem{PRB99075130} H. Wang, J. Ruan, and H. Zhang, Non-Hermitian nodal-line semimetals with an 
anomalous bulk-boundary correspondence, Phys. Rev. B \textbf{99}, 075130 (2019).

\bibitem{PRL118045701} Y. Xu, S.-T. Wang, and L.-M. Duan, Weyl Exceptional Rings in a Three-Dimensional 
Dissipative Cold Atomic Gas, Phys. Rev. Lett. \textbf{118}, 045701 (2017).

\bibitem{PRB99041406} J. C. Budich, J. Carlstr\"{o}m, F. K. Kunst, and E. J. Bergholtz, Symmetry-protected 
nodal phases in non-Hermitian systems, Phys. Rev. B \textbf{99}, 041406 (2019).

\bibitem{arxiv181006549} H. Zhou, J. Y. Lee, S. Liu, and B. Zhen, Exceptional Surfaces in PT-Symmetric 
Photonic Systems, arXiv:1810.06549 (2018).

\bibitem{PRB98165148} K. Kawabata, K. Shiozaki, and M. Ueda, Anomalous helical edge states in a 
non-Hermitian Chern insulator, Phys. Rev. B  \textbf{98}, 165148  (2018).

\bibitem{PRL115040402} J. M. Zeuner, M. C. Rechtsman, Y. Plotnik, Y. Lumer, S. Nolte, M. S. Rudner, 
M. Segev, and A. Szameit, Observation of a Topological Transition in the Bulk of a Non-Hermitian System,
Phys. Rev. Lett. \textbf{115}, 040402 (2015).

\bibitem{NM16433} S. Weimann, M. Kremer, Y. Plotnik, Y. Lumer, S. Nolte, K. G. Makris, M. Segev, 
M. C. Rechtsman, and A. Szameit, Topologically protected bound states in photonic parity–time-symmetric 
crystals, Nat. Mater. \textbf{16}, 433 (2017).

\bibitem{Nature565622} Y. H. Yang, Z. Gao, H. R. Xue, L. Zhang, M. J. He, Z. J. Yang, R. J. Singh, Y. D. Chong,
B. L. Zhang, and H. S. Chen, Realization of a three-dimensional photonic topological insulator, Nature \textbf{565},
622 (2019).
\bibitem{PRB84205128} K. Esaki, M. Sato, K. Hasebe, and M. Kohmoto, Edge states and topological 
phases in non-Hermitian systems, Phys. Rev. B \textbf{84}, 205128 (2011).

\bibitem{PRL102065703} M. S. Rudner and L. S. Levitov, Topological Transition in a Non-Hermitian 
Quantum Walk, Phys. Rev. Lett. \textbf{102}, 065703 (2009).

\bibitem{NP7971} S. Diehl, E. Rico, M. A. Baranov, and P. Zoller, Topology by dissipation in atomic quantum 
wires, Nat. Phys. \textbf{7}, 971 (2011).

\bibitem{PRL832636} M. Creutz, End States, Ladder Compounds, and Domain-Wall Fermions,
 Phys. Rev. Lett. \textbf{83}, 2636 (1999).

\bibitem{JOSA55261} A. L. Jones, Coupling of optical fibers and scattering in fibers, JOSA, \textbf{55}, 261 
(1965) .

\bibitem{OE2006146055}  A. Szameit, J. Burghoff, T. Pertsch, S. Nolte, A. T\"nnermann, and F. Lederer, 
Two-dimensional soliton in cubic fs laser written waveguide arrays in fused silica, Optics Express \textbf{14}, 
6055 (2006).

\bibitem{NN12675} Z. Li, M.H. Kim, C. Wang, Z. Han, S. Shrestha, A.C. Overvig, M. Lu, A. Stein, 
A.M. Agarwal, M. Loncar and N. Yu, Controlling propagation and coupling of waveguide modes using 
phase-gradient metasurfaces, Nature Nanotechnology \textbf{12}, 675 (2017).

\bibitem{NP7907} M. Hafezi, E. A. Demler, M. D. Lukin, and J. M. Taylor,  Robust optical delay lines with 
topological protection, Nat. Phys. \textbf{7}, 907 (2011).
 
\bibitem{MC} G. Golub and C. F. van Loan, Matrix Computations (Johns Hopkins University Press, Baltimore, 2013).
\bibitem{SB631385} L. Zhang, S. Niu, and X.-J. Liu, Dynamical classification of topological quantum phases, 
Science Bulletin \textbf{63}, 1385 (2018).

\bibitem{Sachdev} S. Sachdev, Quantum Phase Transitions (Cambridge University Press, Cambridge, 
England, 1999) .

\bibitem{PRB92085118} L. Li and S. Chen, Characterization of topological phase transitions via topological 
 properties of transition points, Phys. Rev. B \textbf{92}, 085118 (2015).
 \end{thebibliography}
\end{document}